# Anomalous Size Dependence of Relaxational Processes

Armin Bunde,[1,2] Shlomo Havlin,[1,2] Joseph Klafter,[3] Gernot Gräff,[1] and Arkady Shehter[2]

[1]*Institut für Theoretische Physik III, Justus-Liebig-Universität Giessen, D-35392 Giessen, Germany*
[2]*Minerva Center and Department of Physics, Bar-Ilan University, Ramat Gan, Israel*
[3]*School of Chemistry, Tel Aviv University, Tel Aviv, Israel*
(Received 3 December 1996)

We consider relaxation processes that exhibit a stretched exponential behavior. We find that in those systems, where the relaxation arises from two competing exponential processes, the size of the system may play a dominant role. Above a crossover time $t_\times$ that depends *logarithmically* on the size of the system, the relaxation changes from a stretched exponential to a simple exponential decay, where the decay rate also depends *logarithmically* on the size of the system. This result is relevant to large-scale Monte Carlo simulations and should be amenable to experimental verification in low-dimensional and mesoscopic systems. [S0031-9007(97)02972-4]

In recent years it has become clear that many relaxational processes in macroscopic systems can be characterized by a relaxation function $Q(t)$ that exhibits a stretched exponential behavior

$$Q(t) \sim Q(0) \exp[-(t/\tau)^\beta], \qquad (1)$$

where $0 < \beta < 1$. Examples include viscoelastic relaxation [1], dielectric relaxation [2], glassy relaxations [3–5], relaxation in polymers [6,7], and long-time decay in trapping processes [8]. Many more examples [9–13] suggest that Eq. (1) is common to a very wide range of phenomena and macroscopic materials.

The origin of the stretched exponential is not always clear. In many systems it is assumed to be the result of a competition between two exponential processes. In some cases, e.g., trapping processes at long times, this assumption is well established, while in others, such as relaxation in glassy materials, this assumption has been controversially discussed [14,15] and alternative models have been also suggested [10,16–18]. Less is known, both experimentally and theoretically, on the corresponding behavior in mesoscopic systems where we expect the relaxation to depend on the system size.

In this Letter we argue that if the stretched exponential is due to two competing exponential processes, there exists a characteristic time $t_\times$, which depends logarithmically on the size of the system, above which there is a crossover to an exponential decay. Thus, by varying the size of the system this crossover time changes. This can serve as an experimental test for identifying the origin of the mechanism leading to stretched exponential decay.

We assume that the relaxation function of the whole system can be represented by an integration over all possible states $n$, namely,

$$Q(t) = \int_0^\infty \Phi(n) Q(n, t) \, dn. \qquad (2)$$

Here, $\Phi(n)$ is the probability that state $n$ is occupied and $Q(n, t)$ is the dynamic relaxation of the $n$th state.

Usually, in the case of a stretched exponential behavior, $\Phi(n)$ is assumed to behave as $\Phi(n) \sim \exp(-an^\alpha)$, while $Q(n, t)$ decays exponentially with time as $Q(n, t) \sim \exp(-bt/n^\gamma)$. A number of dynamical models that yield a stretched exponential decay can be formulated in terms of Eq. (2). These include the long-time behavior in the trapping problem [8], the target problem [18], direct energy transfer [18], hierarchically constrained dynamics [14] and others. We now concentrate on two examples: The first example is a particle diffusing in a $d$-dimensional system with randomly distributed traps, where we are interested in the survival probability $Q(t)$ of a particle. Here the state $n$ represents a particle in a trap-free region of linear size $n$; $\Phi(n)$ is the probability for the occurrence of a size $n$ trap-free region, and $Q(n, t)$ is the survival probability of the particle in this region [8]. The exponent $\alpha$ is the dimension $d$ of the system, and $\gamma = 2$ due to the diffusional motion. The second example is hierarchically constrained dynamics, a model that has been proposed to account for glassy relaxation [14]. This model assumes that the relaxation of level $n$ populated by spins occurs in stages, and the constraint imposed by a faster degree of freedom must relax before a slower degree of freedom can relax. This implies that the time scale of relaxation in one level is subordinated to the relaxation below. A possible realization considered in [14] and here is a system with a discrete series of levels where the relaxation time of level $n$ is $\tau_n \sim n^\gamma$ [corresponding to the exponential form of $Q(n, t)$ in Eq. (2)], and the weight factor of level $n$ is $\Phi(n) \sim e^{-an}$ [12], corresponding to $\alpha = 1$. The first exponential in Eq. (2) is, accordingly, the probability to occupy level $n$ and the second exponential represents the decay of that level.

We can evaluate the long-time behavior of the integral in Eq. (2) using the method of steepest descent. The main contribution to the integral arises from the maximum of the integrand in (2), which is obtained from the minimum of the function $-an^\alpha - bt/n^\gamma$ appearing in the exponent. This yields that the main contribution to

(2) comes from

$$n^* \cong (\gamma bt/\alpha a)^{1/(\alpha+\gamma)},\qquad(3)$$

leading to Eq. (1) with $\beta = \alpha/(\alpha + \gamma) < 1$, and $\tau = (\alpha/b\gamma)a^{-\gamma/\alpha}[\gamma/(\gamma + \alpha)]^{1+\gamma/\alpha}$.

However, as we show below, these arguments are valid only in the thermodynamic limit where the system size is infinite. For a *finite* system with a finite number $N$ of spins (in the hierarchical constraint system) or a finite number $N$ of traps (in the trapping system), the relaxation function depends explicitly on $N$. Since our discussion is quite general for systems described by Eq. (2), in what follows we refer to spins and traps in the above examples as elements.

For a single finite system consisting of $N$ elements, the relaxation function $Q(t)$ represents an average quantity over the $N$ elements

$$Q(t) = \frac{1}{N}\sum_{\{n\}} m(n)Q(n,t),\qquad(4)$$

where the sum is over all possible states $n$ and $m(n)$ is the number of elements at state $n$, with $\sum_{\{n\}} m(n) = N$. Since the sum in (4) is over exponential functions, the value of $Q(t)$ will fluctuate for different sets of $N$. There will be a distribution of $Q(t)$, and we are interested in the typical $Q(t)$, which is around the peak of this distribution.

In the thermodynamic limit $N \longrightarrow \infty$, all states $n$ are occupied, $m(n)/N$ can be identified with $\Phi(n)$, and Eq. (2) follows. For $N$ finite, in contrast, there exists a characteristic "maximum" state $n = n_{\max}(N)$, and this $n_{\max}$ should replace the upper limit ($\infty$) in Eq. (2),

$$Q(t) = \int_0^{n_{\max}} \Phi(n)Q(n,t)\,dn.\qquad(5)$$

To estimate how $n_{\max}$ depends on $N$, we note that the typical number of states $n$ in a sample of $N$ elements is $Z(n) \cong N\Phi(n) \cong N\exp(-an^\alpha)$. States with $Z(n) \ll 1$ will not occur in a typical system of $N$ elements, and this yields

$$n_{\max} \cong \left(\frac{\ln N}{a}\right)^{1/\alpha}.\qquad(6)$$

If $n^* \ll n_{\max}$, the upper limit in (2) can be approximated by infinity and thus leads to Eq. (1). However, if $n^* \gg n_{\max}$ the main contribution to Eq. (5) will not be from the maximum of the integrand, which is outside the range of integration, but from $n_{\max}$. Thus, for $n^* \gg n_{\max}$ we expect

$$Q(t) \cong Q(0)e^{-bt/n_{\max}^\gamma},\qquad(7)$$

where the time constant of the relaxation, $n_{\max}^\gamma$, scales as $(\ln N)^{\gamma/\alpha}$. The crossover time from a stretched exponen-tial [Eq. (1)] to an exponential [Eq. (7)] can be estimated from the condition $n^* = n_{\max}$, from which follows

$$t_\times \cong \frac{\alpha a}{\gamma b}\left(\frac{\ln N}{a}\right)^{1+\gamma/\alpha}.\qquad(8)$$

The striking point in Eq. (8) is the logarithmic dependence on $N$, which puts $t_\times$ in the range of observable time scales measurable in mesoscopic systems. Indeed, the corresponding relaxation value $Q(t_\times)$ scales as

$$Q(t_\times) \sim N^{-\alpha/\gamma},\qquad(9)$$

independent of the microscopic parameters $a$ and $b$. In the case of the trapping relaxation mechanism where $\alpha = d$ and $\gamma = 2$ we obtain

$$Q(t_\times)/Q(0) \sim N^{-d/2},\qquad(10)$$

$Q(t_\times)/Q(0) \sim N^{-d/2}$, while in the hierarchical constraint dynamics

$$Q(t_\times)/Q(0) \sim N^{-1/\gamma}.\qquad(11)$$

It is known [8(e)] that in both examples, for an infinite system, the stretched exponential behavior of Eq. (1) sets in only at very long times. Thus we expect that in the finite system, the crossover will mask the stretched-exponential pattern.

To test our analytical approach, we performed Monte Carlo simulations on both the trapping model and the hierarchical constraint model. In the trapping model, we consider one- and two-dimensional systems with a fixed concentration $c = 0.5$ of randomly distributed traps and vary the size $N/c$ of the system. We calculated numerically the survival probability $Q(t)$ of a particle as a function of $t$ and $N$. In the hierarchical model we have chosen $\tau_n \sim n$, i.e., $\gamma = 1$. We calculated the relaxation function for system sizes varying from $N = 10^2$ to $N = 10^5$.

As mentioned earlier, the relaxation function fluctuates for different sets of $N$. For obtaining the typical behavior of $Q(t)$, we have considered therefore the "typical" average $Q(t)_{\text{typ}} \equiv \exp(\langle \ln Q(t)\rangle)$, where the brackets denote an average over many sets of $N$ elements [19]. For simplicity, we shall drop the index "typ" in the following.

Figure 1 shows $-\ln[Q(t)/Q(0)]$ as a function of $t$ in a double logarithmic plot for (a) the trapping model in $d = 1$ and $d = 2$, and (b) the hierarchical constraint model, both for several system sizes. In all cases, a crossover from an exponent $\beta < 1$ (at small $t$) towards $\beta = 1$ (at large $t$) can be easily recognized. The crossover time $t_\times$ shifts towards larger values when $N$ increases.

To study the crossover behavior in a more quantitative manner, we have plotted in Fig. 2 the local exponents $\beta$

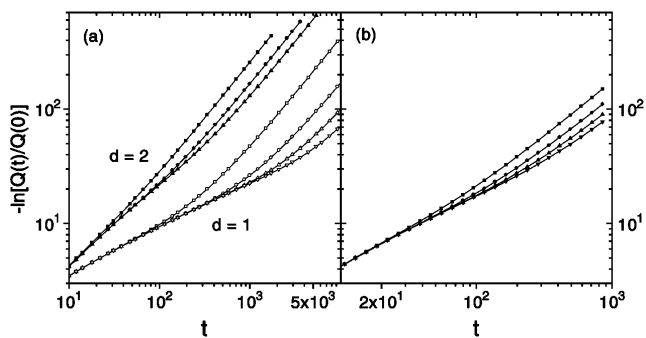

FIG. 1. Plot of $-\ln[Q(t)/Q(0)]$ as a function of $t$ in a double logarithmic presentation for (a) the trapping model in $d = 1$ and $d = 2$, and (b) the hierarchical constraint model, for several system sizes. For the trapping model, the system sizes are $N = 2 \times 10^3$ (open square), $2 \times 10^5$ (open circle), $2 \times 10^7$ (open up triangle), $2 \times 10^9$ (open down triangle) in $d = 1$, and $N = 9 \times 10^2$ (full square), $9 \times 10^4$ (full circle), $9 \times 10^6$ (full up triangle) in $d = 2$. For the hierarchical model, the system sizes are $N = 10^2$ (full square), $10^3$ (full circle), $10^4$ (full up triangle), $10^5$ (full down triangle).

obtained from the local slopes of Fig. 1, as a function of $t$. In both systems, for a fixed system size $N$, $\beta$ first decreases with $t$, reaches a minimum value at a certain time that can be identified with $t_\times$, and then increases monotonically with time towards $\beta = 1$. The figure shows that the minimum value of $\beta$ has not yet reached its asymptotic value predicted for infinite systems, i.e., $\beta = 1/3$ ($d = 1$) and $\beta = 1/2$ ($d = 2$) for the trapping system and $\beta = 1/2$ for the hierarchical system.

To show the dependence of the crossover time $t_\times$ on the system size $N$ we have plotted, in Fig. 3, the values of $t_\times^{\alpha/(\alpha+\gamma)}$ as a function of $\ln N$. The crossover time was obtained numerically from the position of the minima of the curves in Fig. 2. The resulting straight lines are in full agreement with the prediction of Eq. (8), supporting our analytical approach.

In the following we discuss the relevance of our results to Monte Carlo simulations and experiments. There exists a long standing puzzle in Monte Carlo simulations of the trapping problem in $d = 2$ and 3, that the predicted stretched exponential could not be observed [8], even for survival probabilities $Q(t)/Q(0)$ down to $10^{-21}$ in $d = 2$ [8(b)] and $10^{-67}$ in $d = 3$ [8(g)].

Our finding of the logarithmic dependence of $Q(t)$ on the system size $N$ explains this puzzle. The Monte Carlo simulations in $d = 2$ and 3 were typically performed on $10^3$ configurations with about $10^4$ traps, which is equivalent to having a single system with $N \sim 10^7$ traps. Using Eq. (10), we expect for $N = 10^7$ traps $Q(t_\times)/Q(0) \cong 10^{-7}$ in $d = 2$. Indeed, for times above $t_\times$ the exponent $\beta$ approaches unity as predicted by our theory and as seen clearly in Fig. 2(a). Moreover, for this system size $\beta$ never reaches the predicted thermodynamic value $\beta = 0.5$, the minimum value of $\beta$ is about 0.65. For $d = 3$, $Q(t_\times)/Q(0) \cong 10^{-11}$ thus for smaller survival values ($t > t_\times$) one again expects increasing values of $\beta$ approaching unity. This explains the exponential decay found in the early Monte Carlo simulations. Our results show that this is not an artifact but due to the finite size of the system. Moreover, they clearly indicate that the thermodynamic limit cannot even be reached in one-dimensional macroscopic systems.

It would be of interest to test the above prediction experimentally by preparing experimental realizations where size effects can be controlled. Equations (8) and (10) suggest that the behavior around the crossover can be measured experimentally. For the trapping problem in linear systems, which has been studied experimentally [20,21], we expect for $10^8$ sites and concentrations of traps $c$ between $10^{-4}$ and $10^{-2}$, that $Q(t_\times)/Q(0) \sim 10^{-2}–10^{-3}$, which is a survival range that can be detected experimentally. The same arguments are valid for the target problem and therefore a similar crossover from stretched exponential to exponential decay is expected in relaxation experiments in low-dimensional geometries [22]. Mesoscopic systems such as quantum dots, are also promising candidates for experiments where the crossover can be relevant. Identifying the logarithmic size dependence in experiments may provide support to the theories claiming that the observed stretched exponential is due to competing exponential processes, represented by Eq. (2).

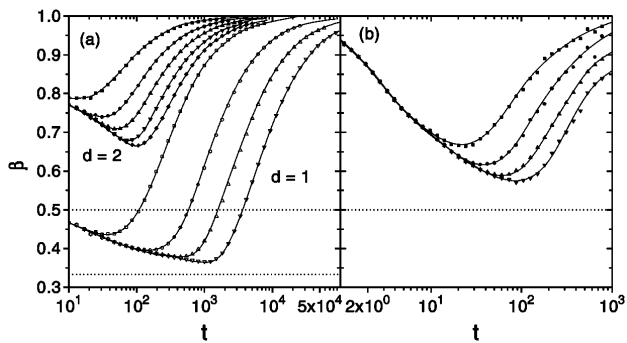

FIG. 2. Plot of the local exponents $\beta$ calculated from the successive slopes of the corresponding curves in Figs. 1(a), for the trapping model and 1(b) for the hierarchical model. The horizontal dashed lines represent the corresponding asymptotic ($N \longrightarrow \infty$, $t \longrightarrow \infty$) values of $\beta$.

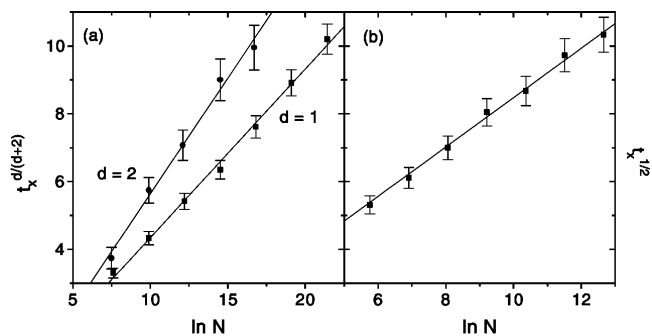

FIG. 3. Plot of $t_\times^{\alpha/(\alpha+\gamma)}$ as a function of $\ln N$, for (a) the trapping model and (b) the hierarchical model. The straight line supports Eq. (8). The crossover times $t_\times$ were obtained from the positions of the minima of Fig. 2.

We would like to thank Julia Dräger for valuable discussions. This work was supported by the German Israeli Foundation (GIF).

———————

[1] R. Kohlrausch, Ann Phys. (Leipzig) **12**, 393 (1847).

[2] G. Williams and D.C. Watts, Trans. Faraday Soc. **66**, 80 (1970).

[3] V. Chamberlin, G. Mozurkewich, and R. Orbach, Phys. Rev. Lett. **52**, 867 (1984)

[4] F. Mezei and A.P. Murani, J. Magn. Magn. Mater. **14**, 211 (1979).

[5] A. Plonka, *The Dependent Reactivity of Species in Condensed Matter* (Springer-Verlag, New York, 1986).

[6] A.A. Jones *et al.*, Macromolecules **16**, 658 (1983).

[7] K.L. Li *et al.*, Macromolecules **21**, 2940 (1983).

[8] (a) N.D. Donsker and S.R.S. Varadhan, Commun. Pure Appl. Math. **32**, 721 (1979); (b) P. Grassberger and I. Procaccia, J. Chem. Phys. **77**, 6281 (1982); (c) J. Klafter, G. Zumofen, and A. Blumen, J. Phys. Lett. **45**, L49 (1984); (d) I. Webman, Phys. Rev. Lett. **52**, 220 (1984); (e) S. Havlin, M. Dishon, J.E. Kiefer, and G.H. Weiss, Phys. Rev. Lett. **53**, 407 (1984); (f) M. Fixman, Phys. Rev. Lett. **52**, 791 (1984); (g) J.K. Anlauf, Phys. Rev. Lett. **52**, 1845 (1984).

[9] A.K. Jonscher, Nature (London) **267**, 673 (1977).

[10] K.L. Ngai, Comments Solid State Phys. **9**, 127 (1979); **9**, 141 (1980).

[11] K. Funke, Prog. Solid State Chem. **22**, 11 (1993).

[12] J. Klafter and M.F. Shlesinger, Proc. Natl. Acad. Sci. U.S.A. **83**, 848 (1986).

[13] H. Scher, M.F. Shlesinger, and J.T. Bendler, Phys. Today **44**, No. 1, 26 (1991).

[14] R.G. Palmer, D.L. Stein, E. Abrahams, and P.W. Anderson, Phys. Rev. Lett. **53**, 958 (1984).

[15] W. Götze and L. Sjögren, Rep. Prog. Phys. **55**, 241 (1992).

[16] M.H. Cohen and G.S. Grest, Phys. Rev. B **24**, 4091 (1981).

[17] M.F. Shlesinger and E.W. Montroll, Proc. Natl. Acad. Sci. U.S.A. **81**, 1280 (1984).

[18] A. Blumen, J. Klafter, and G. Zumofen, in *Optical Spectroscopy of Glasses,* edited by I. Zchokke (Reidel, Dordrecht, 1986).

[19] An arithmetic average over $M$ sets of $N$ elements cannot be employed here, since it leads to a result identical for a larger system with $M \times N$ elements [see Eq. (4)]. For a discussion of the typical average in the context of random walks on random fractals (where $N$ represents the number of random configurations taken into account in the average and a $\ln N$ dependence occurs in the probability density), see A. Bunde and J. Dräger, Phys. Rev. E **52**, 53 (1995); J. Dräger and A. Bunde *ibid.* **54**, 4596 (1996).

[20] R.A. Auerbach and G.L. McPherson, Phys. Rev. B **33**, 6815 (1986).

[21] R. Knockenmuss and H.U. Gudel, J. Chem. Phys. **86**, 1104 (1987).

[22] J.M. Drake *et al.*, Phys. Rev. Lett. **61**, 865 (1988).